# A BAYESIAN APPROACH TO SHAPE RECONSTRUCTION OF A COMPACT OBJECT FROM A FEW NUMBER OF PROJECTIONS

ALI MOHAMMAD-DJAFARI

*Laboratoire des Signaux et Systèmes (CNRS-ESE-UPS)*
*École Supérieure d'Électricité,*
*Plateau de Moulon, 91192 Gif-sur-Yvette, France.*
*E-mail: djafari@lss.supelec.fr*

**Abstract.** Image reconstruction in X ray tomography consists in determining an object from its projections. In many applications such as non destructive testing, we look for an image who has a constant value inside a region (default) and another constant value outside that region (homogeneous region surrounding the default). The image reconstruction problem becomes then the determination of the shape of that region. In this work we model the object (the default region) as a polygonal disc and propose a new method for the estimation of the coordinates of its vertices directly from a very limited number of its projections.

## 1. Introduction

Gammagraphy is a well known technique in non destructive testing (NDT) and non destructive evaluation (NDE) applications. Tomographic image reconstruction in these applications is more recent and consists of determining an object from its projections. The relation between the object $f(x, y)$ and its projections $p(r, \phi)$ is frequently modeled by the Radon transform:

$$p(r, \phi) = \iint f(x, y) \delta(r - x \cos \phi - y \sin \phi) \, dx \, dy \qquad (1)$$

In many image reconstruction applications, especially in NDT and NDE, we look for an image $f(x, y)$ who has a constant value $c_1$ inside a region (default region D) and another constant value $c_2$ outside that region (homogeneous surrounding safe region), e.g. metal & air. The image reconstruction problem becomes then the determination of the shape of the default region.

In this communication, without loss of generality, we assume that $c_1 = 1$ and $c_2 = 0$:

$$f(x, y) = \begin{cases} 1 & \text{if} \quad (x, y) \in D, \\ 0 & \text{elsewhere} \end{cases}, \qquad (2)$$

where $D$ represents the default region.

There has been many works in image reconstruction and computed tomography dealing with this problem. To emphasis the originality and the place of this work, we give here a summary of the different approaches for this problem:

• In the first approach, one starts by discretizing the equation (1) to obtain:

$$\boldsymbol{p} = \boldsymbol{H}\boldsymbol{f} + \boldsymbol{n} \tag{3}$$

where, $\boldsymbol{f}$ is the discretized values of the object $f(x, y)$ (the pixel values of the image), $\boldsymbol{p}$ is values of the projection data $p(r, \phi)$, $\boldsymbol{n}$ is a vector to represent the modeling and measurement errors (noise) and $\boldsymbol{H}$ the discretized Radon operator. Then the solution is defined as the argument which minimizes the regularization criterion

$$J(\boldsymbol{f}) = ||\boldsymbol{p} - \boldsymbol{H}\boldsymbol{f}||^2 + \lambda\Omega(\boldsymbol{f}), \tag{4}$$

where $\lambda$ is the regularization parameter.

$\Omega(\boldsymbol{f})$ has to be chosen appropriately to reflect the fact that $\boldsymbol{f}$ must represent a binary image. This is the classical approach of general image reconstruction problem. In fact, one can also interpret $J(\boldsymbol{f})$ as the maximum a posteriori (MAP) criterion in the Bayesian estimation framework where $Q(\boldsymbol{f}) = ||\boldsymbol{p} - \boldsymbol{H}\boldsymbol{f}||^2$ represents the likelihood term and $\exp\left[-\lambda\Omega(\boldsymbol{f})\right]$ the prior probability law.

This approach has been used with success in many applications (e.g. [1, 2, 3, 4]) but the cost of its calculation is huge due to the great dimension of $\boldsymbol{f}$. Many works have been done on choosing appropriate regularization functionals or equivalently appropriate prior probability laws for $\boldsymbol{f}$ to enforce some special properties of the image such as smoothness, positivity or piecewise smoothness [5, 6, 7, 8, 9, 10]. Among these, one can mention mainly two types of functions for $\Omega(\boldsymbol{f})$:

Entropic laws:

$$\Omega(\boldsymbol{f}) = \sum_{j=1}^{N} \phi(f_j), \quad \text{with} \quad \phi(x) = \{x^p, -x\log x, \log x, \cdots\}$$

Homogeneous Markovian laws:

$$\sum_{j=1}^{N} \sum_{i \in \mathcal{N}_j} \phi(f_j, f_i), \quad \text{with } \phi(x, y) = \left\{|x-y|^p, -|x-y|\log\frac{x}{y}, \log\cosh|x-y|, \cdots\right\}$$

See for example [6] for the entropic laws, [7, 5] for scale invariant markovian laws and [8, 9, 10] for other specific choices.

• In the second approach, one starts by giving a parametric model for the object and then tries to estimate these parameters using least squares (LS) or maximum likelihood (ML) methods. In general, in this approach one chooses a parametric model such as superposition of circular or elliptical discs to be able to relate analytically the projections to these parameters. For example, for a superposition of elliptical discs we have:

$$f(x, y) = \sum_{k=1}^{K} d_k f_k(x, y) \tag{5}$$

with

$$f_k(x,y) = \begin{cases} 1 & \text{if } (x - \alpha_k)^2 + (y - \beta_k)^2 < g_k^2(\theta), \\ 0 & \text{elsewhere} \end{cases}, \tag{6}$$

where

$$g_k(\theta) = \sqrt{a_k^2 \cos^2 \theta + b_k^2 \sin^2 \theta}. \tag{7}$$

and where $\boldsymbol{\theta} = \{d_k, \alpha_k, \beta_k, a_k, b_k, k = 1, \cdots, K\}$ is a vector of parameters defining the parametric model of the image (density values, coordinates of the centers and the two diameters of the ellipses). It is then easy to calculate analytically the projections and the relation between the data and the unknown parameters becomes:

$$p(r, \phi) = h(r, \phi; \boldsymbol{\theta}) + n(r, \phi) \tag{8}$$

where

$$h(r, \phi; \boldsymbol{\theta}) = \sum_{k=1}^{K} d_k p_k(r, \phi) \quad \text{with} \quad p_k(r, \phi) = \begin{cases} \frac{2a_k b_k}{g_k^2(\phi)} \sqrt{g_k^2(\phi) - r^2)} & \text{if } r < g_k(\phi), \\ 0 & \text{elsewhere} \end{cases} \tag{9}$$

The LS or the ML estimate when the noise is assumed to be Gaussian is then given by:

$$\widehat{\boldsymbol{\theta}} = \arg\min_{\boldsymbol{\theta}} \left\{ \|p(r, \phi) - h(r, \phi; \boldsymbol{\theta})\|^2 \right\} \tag{10}$$

This approach has also been used with success in image reconstruction [11, 12, 13, 14, 15]. But, the range of applicability of these methods is limited to the cases where the parametric models are actually appropriate.

• In the third approach which is more appropriate to our problem of shape reconstruction, one starts by modeling directly the contour of the object by a function, say $g(\theta)$ such as:

$$D = \left\{ (x, y) : \rho^2(\theta) = x^2 + y^2 < g^2(\theta) \right\}. \tag{11}$$

The next step is then to relate the projections $p(r, \phi)$ to $g(\theta)$ which, in this case is:

$$p(r, \phi) = \int_0^{2\pi} \int_0^{g(\theta)} \delta(r - \rho \cos(\phi - \theta)) \rho \, d\rho \, d\theta. \tag{12}$$

and finally to discretize this relation to obtain:

$$\boldsymbol{p} = \boldsymbol{h}(\boldsymbol{g}) + \boldsymbol{n} \tag{13}$$

where $\boldsymbol{g}$ represents the discretized values of $g(\theta)$ defining the contour of the object and $\boldsymbol{h}(\boldsymbol{g})$ represents the discretized version of the nonlinear operator (12) relating projection data $\boldsymbol{p}$ and $\boldsymbol{g}$. Then, one defines the solution as the argument which minimizes

$$J(\boldsymbol{g}) = \|\boldsymbol{p} - \boldsymbol{h}(\boldsymbol{g})\|^2 + \lambda \Omega(\boldsymbol{g}), \tag{14}$$

where $\Omega(\boldsymbol{g})$ has to be chosen appropriately to reflect some regularity property of the object's contour.

In this case also one can consider $J(\boldsymbol{g})$ as the MAP criterion with $Q(\boldsymbol{g}) = ||\boldsymbol{p} - \boldsymbol{h}(\boldsymbol{g})||^2$ as the likelihood term and $\Omega(\boldsymbol{g})$ as the prior one.
This approach has been used in image restoration [16], but it seems to be new in image reconstruction applications and the proposed method in this work is in this category. The originality of our work is to model the contour of the object by a piecewise linear function which means that the object is modeled as a polygonal disc whose vertices are estimated directly from the projection data.

Now before going further in details, let compare this last approach with the first one, by noting the following:

• In (3) and (4), $\boldsymbol{f}$ represents the pixel values of the image (a very great dimensional vector depending on the image dimensions), but in (13) and (14), $\boldsymbol{g}$ represents the discretized values of $g(\theta)$ defining the contour of the object. The dimension of this vector is moderate and independent of the image dimensions.

• In (3) and (4), $\boldsymbol{H}\boldsymbol{f}$ is a linear function of $\boldsymbol{f}$ and so $Q(\boldsymbol{f})$ is a quadratic function of it, but in (13) and (14), $\boldsymbol{h}(\boldsymbol{g})$ is not a linear function of $\boldsymbol{g}$ and so $Q(\boldsymbol{g})$ will not be a quadratic function of it.
We will discuss more the consequences of these remarks in the next section.

## 2.  Proposed method

In this paper we propose to model the contour of the object (default region) as a periodic piecewise linear function or equivalently to model the shape of the object as a polygonal disc with a great number $N$ of vertices to be able to approximate any shape. Then we propose to estimate directly the coordinates $\{(x_j, y_j), j = 1, \cdots, N\}$ of the vertices of this polygonal disc from the projection data (see Fig. 2.).

The idea of modeling the shape of the object as a polygonal disc is not new and some works have been done in image reconstruction applications, but, in general in these works, a hypothesis of convexity of the polygonal disc has been used which is very restrictive in real applications. In our work we do not make this hypothesis and also we choose $N$ appropriately great to to be able to approximate any shape.

As we deal with inverse problems, the solution is then defined as the argument which minimizes the following criterion

$$J(\boldsymbol{z}) = ||\boldsymbol{p} - \boldsymbol{h}(\boldsymbol{z})||^2 + \lambda \Omega(\boldsymbol{z}), \tag{15}$$

where $\boldsymbol{z} = \boldsymbol{x} + i\boldsymbol{y}$ is a complex vector whose real and imaginary parts represent the $x$ and the $y$ coordinates of the polygon vertices, $\boldsymbol{h}(\boldsymbol{z})$ represents the direct operator which calculates the projections for any given $\boldsymbol{z}$ and $\Omega(\boldsymbol{z})$ is chosen to be a function which reflects the regularity of the object contour. In this work we used the following:

$$\Omega(\boldsymbol{z}) = \sum_{j=1}^{N} |z_{j-1} - 2z_j + z_{j+1}|^2. \tag{16}$$

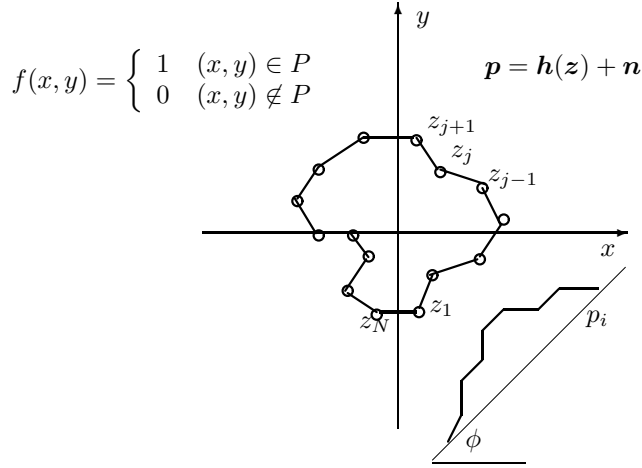

Figure 1: Proposed shape reconstruction modeling.

Note that $|z_{j-1}-2z_j+z_{j+1}|^2$ is just the Euclidian distance between the point $z_j$ and the line segment passing through $z_{j-1}$ and $z_{j+1}$ and so this choice favors a shape whose local curvature is limited. We can also give a probabilistic interpretation to this choice. In fact we can consider $z_j$ as random variables with the following Markovian law:

$$p(z_j|\boldsymbol{z}) = p(z_j|z_{j-1}, z_{j+1}) \propto \exp\left[-\frac{1}{2\sigma^2}|z_{j-1}-2z_j+z_{j+1}|^2\right] \qquad (17)$$

Other functions are possible and are studied in this work.

In both cases, the criterion $J(\boldsymbol{z})$ is multimodal essentially due to the fact that $\boldsymbol{h}(\boldsymbol{z})$ is a nonlinear function of $\boldsymbol{z}$. Calculating the optimal solution corresponding to the global minimum of (15) needs then carefully designed algorithms. For this we propose the following strategies:

• The first is to use a global optimization technique such as simulated annealing (SA). This technique has given satisfactory result as it can be seen from the simulations in the next section. However, this algorithm needs a great number of iterations and some skills for choosing the first temperature and cooling schedule, but the overall calculations is not very important due to the fact that we do not need to calculate the gradient of the criterion (15).

• The second is to find an initial solution in the attractive region of the global optimum and to use a local descent type algorithm to find the solution.

The main problem here is how to find this initial solution. For this, we used a moment based method proposed by Milanfar, Karl & Wilsky [17, 18] which is accurate enough to obtain an initial solution which is not very far from the optimum. The basic idea of this method is to relate the moments of the projections to the moments of a class of polygonal discs obtained by an affine transformation of a centered regular polygonal disc, and so to estimate a polygonal disc whose vertices are on an ellipse and whose moments up to the second order matches those of the projections.

However, there is no theoretical proof that this initial solution will be in the attractive region of the global optimum. In the simulation results section we will show some results comparing the performances of these two methods as well as a comparison with some other classical methods.

## 3.  Simulation results

To measure the performances of the proposed method and keeping the objective of using this method for NDT applications where the number of projections are very limited, we simulated a case where the object is a polygonal disc with $N = 40$ vertices (hand-made) and calculated its projections for only 5 directions:

$$\phi = \{-45, -22.5, 0, +22.5, +45 \text{ degrees}\}$$

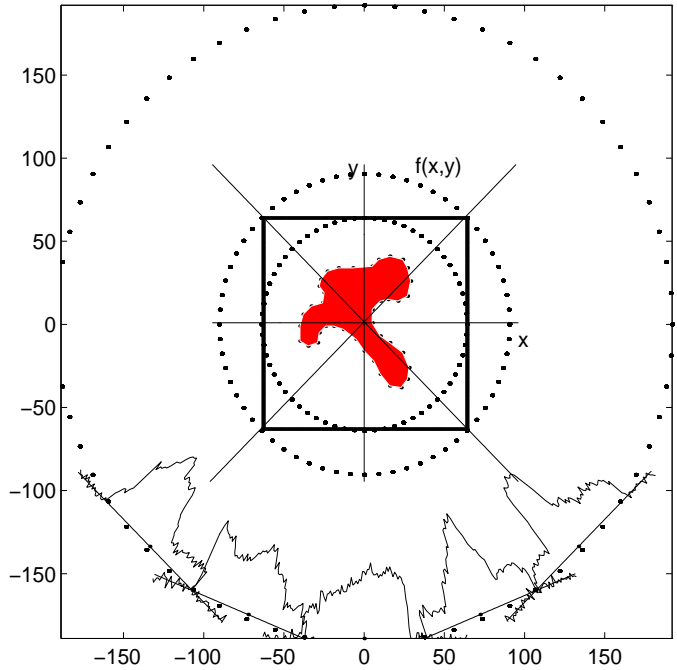

Figure 2: Original image and noisy projections.

Then, we added some noise (white, Gaussian and centered) on them to simulate the measurement errors. The S/N ratio was chosen 20dB. Finally, from these data we estimated the solution by either of the two proposed methods. Figures 3. and 3. show these results.

In Fig. 3., we give the reconstruction results obtained by simulated annealing (SA) algorithm and in Fig. 3. those obtained by a moment-based initialization and a local descent-based optimization algorithm. Note that, the SA is independent of initialization, however, in these figures we show the results obtained by the proposed method.

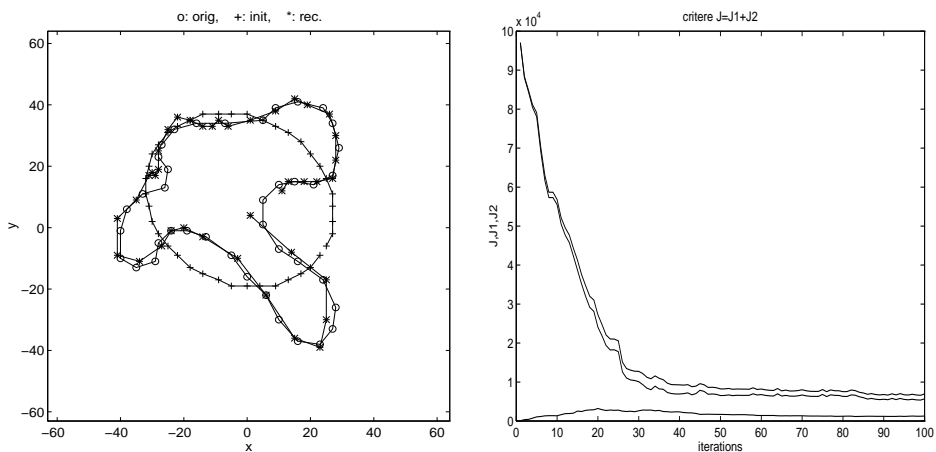

Figure 3:   Reconstruction using simulated annealing.
a) Original, Initialization and Reconstructed objects
b) Evolution of the criterion : $J = J_1 + \lambda J_2$ where $J1 = Q(\boldsymbol{z})$ and $J_2 = \Omega(\boldsymbol{z})$

In Fig. 3. we show a comparison between the results obtained by the proposed method and those obtained either by a classical backprojection method or by some other methods in the first approach using (3) and (4) with different regularization functionals $\Omega(\boldsymbol{f})$ among those in (5). Also, for the purpose of curiosity we show the result of a binary segmented image obtained by thresholding these last images.

## 4.   Conclusions

A new method for tomographic image reconstruction of a compact object from its limited angle projections is proposed. The basic idea of the proposed method is to model the object as a polygonal disc whose vertices coordinates are estimated directly from the projections using the Bayesian MAP estimation framework or equivalently by optimizing a regularized criterion.

This criterion is not unimodal. To optimize it two methods are examined: a global optimization method based on simulated annealing and a local gradient-based method with a good initialization obtained using a moment based method. The first one seems to give entire satisfaction and better results. The final destination of the proposed method is for non destructive testing (NDT) and evaluation (NDE) image reconstruction applications including X-rays, ultrasound or Eddy currents [19, 20, 21].

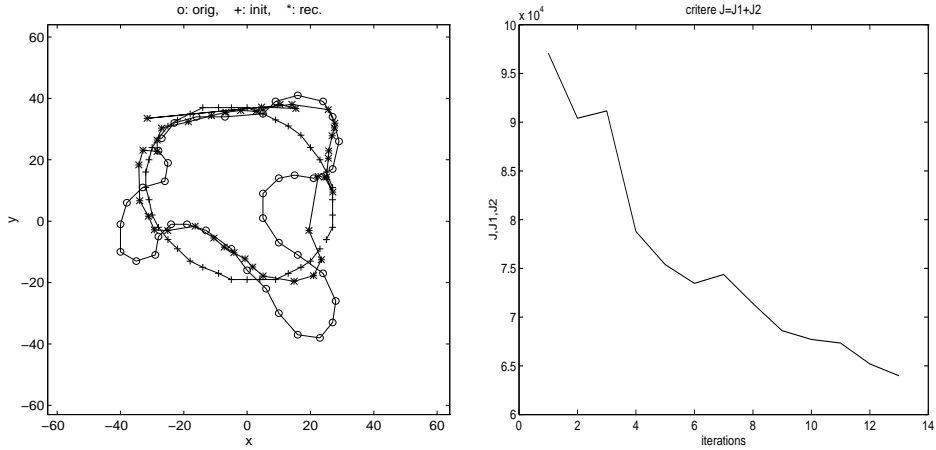

Figure 4: Reconstruction using a moment-based initialization and a local mini-mizer.
a) Original, Initialization and Reconstructed objects
b) Evolution of the criterion $J = J_1 + \lambda J_2$ during the iterations.

## References

1. G. Herman, H. Tuy, H. Langenberg, and P. Sabatier, *Basic Methods of Tomography and Inverse Problems.* Adams Hilger, 1987.

2. A. Kak and M. Slaney, *Principles of Computerized Tomographic Imaging.* New York, NY: IEEE Press, 1987.

3. S. Geman and D. McClure, "Statistical methods for tomographic image reconstruc-tion," in *Proc. of the 46-th Session of the ISI, Bulletin of the ISI*, vol. 52, pp. 22–26, 1987.

4. G. Demoment, "Image reconstruction and restoration : Overview of common es-timation structure and problems," *ieeeASSP*, vol. ASSP-37, pp. 2024–2036, Dec. 1989.

5. S. Brette, J. Idier, and A. Mohammad-Djafari, "Scale invariant Markov models for linear inverse problems," in *Proc. of the Section on Bayesian Statistical Sciences*, (Alicante, Spain), pp. 266–270, American Statistical Association, 1994.

6. A. Mohammad-Djafari and J. Idier, *A scale invariant Bayesian method to solve linear inverse problems*, pp. 121–134. Maximum entropy and Bayesian methods, Santa Barbara, U.S.A.: Kluwer Academic Publ., g. heidbreder ed., 1996.

7. C. Bouman and K. Sauer, "A generalized Gaussian image model for edge-preserving MAP estimation," *IEEE Transactions on Image Processing*, vol. IP-2, pp. 296–310, July 1993.

8. L. Bedini, I. Gerace, and A. Tonazzini, "A deterministic algorithm for reconstructing images with interacting discontinuities," *Computer Vision and Graphics and Image Processing*, vol. 56, pp. 109–123, March 1994. AMD.

9. M. Nikolova, A. Mohammad-Djafari, and J. Idier, "Inversion of large-support ill-conditioned linear operators using a Markov model with a line process," in *ICASSP*, vol. V, (Adelaide, Australia), pp. 357–360, 1994.

10. M. Nikolova, J. Idier, and A. Mohammad-Djafari, "Inversion of large-support ill-

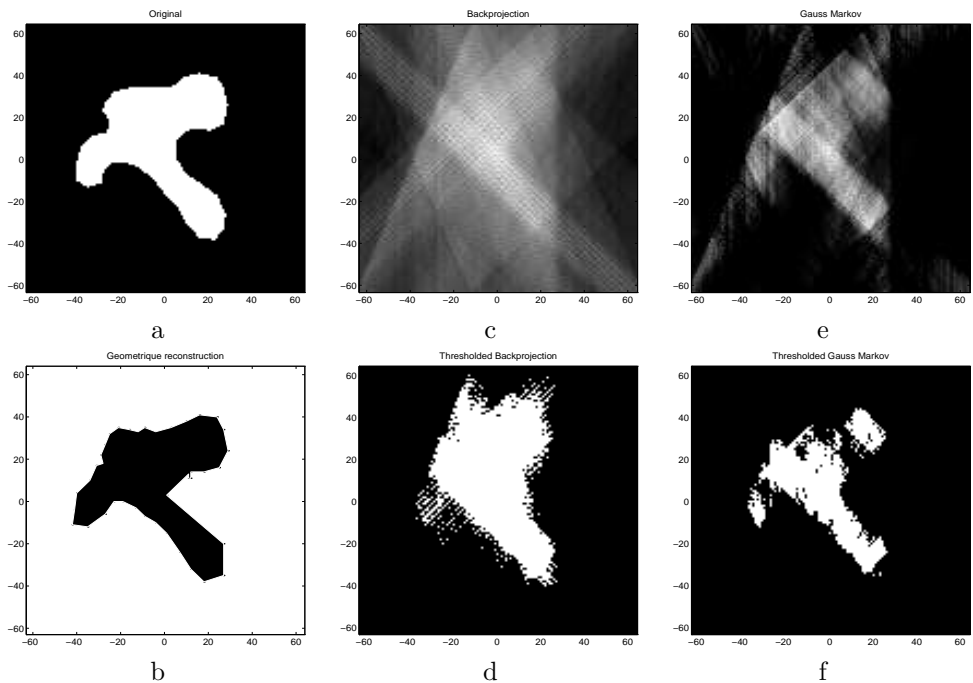

Figure 5:  A comparison with backprojection and some other classical methods
a) Original, b) Proposed method,
c) Backprojection, d) Binary threshold of c),
e) Gaussian Markov Random Field (GMRF) modeling and the MAP estimation
reconstruction, f) Binary threshold of e).

posed linear operators using a piecewise Gaussian MRF," tech. rep., GPI–LSS, submitted to IEEE Transactions on Image Processing, Gif-sur-Yvette, France, 1995.

11. L. A. Shepp and Y. Vardi, "Maximum likelihood reconstruction for emission tomography," *IEEE Transactions on Medical Imaging*, vol. MI-1, pp. 113–122, 1982.

12. A. J. Devaney and G. A. Tsihrintzis, "Maximum likelihood estimation of object location in diffraction tomography," *IEEE Transactions on Signal Processing*, vol. SP-39, pp. 672–681, Mar. 1991.

13. D. J. Rossi and A. S. Wilsky, "Reconstruction from projections based on detection and estimation of objects," *ieeeASSP*, vol. ASSP-32, no. 4, pp. 886–906, 1984.

14. J. L. Prince and A. S. Wilsky, "Reconstructing convex sets from support line measurements," *ieeePAMI*, vol. 12, no. 3, pp. 377–389, 1990.

15. J. L. Prince and A. S. Wilsky, "Convex set reconstruction using prior shape information," *CVGIP*, vol. 53, no. 5, pp. 413–427, 1991.

16. N. S. Friedland and A. Rosenfeld, "Compact object recognition using energy-function-based optimization," *IEEE Transactions on Pattern Analysis and Machine Intelligence*, vol. 14, no. 7, pp. 770–777, 1992.

17. P. Milanfar, *Geometric Estimation and Reconstruction from Tomographic Data*. PhD thesis, MIT, Dept. of Electrical Eng., 1993.

18. P. Milanfar, W. C. Karl, and A. S. Wilsky, "A moment-based variational approach to tomographic reconstruction," *IEEE Transactions on Image Processing*, vol. 25, no. 9, pp. 772–781, 1994.

19. S. Gautier, G. Le Besnerais, A. Mohammad-Djafari, and B. Lavayssière, *Data fusion in the field of non destructive testing*. Maximum entropy and Bayesian methods, Santa Fe, U.S.A.: Kluwer Academic Publ., K. Hanson ed., 1995.

20. D. Prémel and A. Mohammad-Djafari, "Eddy current tomography in cylindrical geometry," *ieeeM*, vol. M-31, pp. 2000–2003, May 1995.

21. M. Nikolova and A. Mohammad-Djafari, "Eddy current tomography using a binary Markov model," *To appear in Signal Processing*, vol. 49, pp. 000–000, May 1996.